\documentclass[11pt]{article}

\usepackage[margin=1in]{geometry}    
\usepackage{graphicx}                
\usepackage{amsmath,amssymb}         
\usepackage{hyperref}                
\usepackage{xcolor}                  
\usepackage{listings}                
\usepackage{caption}                 

\title{\textbf{A Python Toolkit for Plotting Double Star Observations with 1:1 Aspect Ratio}\\
       \large }

\author{
  \textbf{Xinyue Wang} \\
  Stanford University Online High School, Stanford, CA \\
  \texttt{wxinyue@ohs.stanford.edu}
}

\date{January 30, 2025}

\begin{document}

\maketitle

\begin{abstract}
Accurate visualization of double star astrometric data is essential for effective analysis and interpretation. This article presents a Python toolkit designed for astronomers who need to plot measurements from diverse sources---historical, Gaia DR3, and the Las Cumbres Observatory (LCO) network---while maintaining a 1:1 aspect ratio to avoid visually distorting the data. The toolkit is composed of three scripts: one that handles polar coordinates (P.A., separation), one for Cartesian (X, Y) coordinates, and another with the option to include predicted theoretical points. This paper describes the purpose, functionality, and usage of these scripts, including example figures, installation guides, and licensing information.

This toolkit has been used by the author and collaborators in published and submitted research on double star systems, demonstrating its versatility for both professional and student-driven investigations.
\end{abstract}

\tableofcontents

\section{Introduction}

Double stars (or binary stars) are pairs of stars that appear close to each other 
in the night sky. Through long-term observation, astronomers can determine whether 
these pairs are gravitationally bound (binaries), or simply stellar alignments 
(visual doubles). Astrometric measurements often include position angle (P.A.) and 
separation (Sep) in polar coordinates, or Cartesian coordinates (X, Y) derived from 
the polar form.

Over time, large datasets have been accumulated from historical catalogs, modern 
surveys such as Gaia DR3 \cite{GaiaDR3}, and current observation networks like the 
Las Cumbres Observatory (LCO) \cite{LCO}. Combining these measurements allows 
astronomers to trace orbital motion or identify systematic biases. However, 
plotting these data accurately can be challenging, especially if the displayed 
aspect ratio is not carefully controlled. A non-square plot can visually distort 
the geometry and lead to incorrect qualitative conclusions.

In a research seminar at Stanford University Online High School (SUOHS), I developed 
a Python-based toolkit to streamline these plots. Specifically, the scripts 
described here:

\begin{itemize}
  \item Combine historical measurements, Gaia DR3 data, LCO data, and optional predicted points.
  \item Enforce a 1:1 aspect ratio to preserve angular and spatial relationships.
  \item Provide color-coding or custom markers for different data sources.
  \item Offer both polar- and Cartesian-based approaches.
\end{itemize}

The goal is to give astronomers, students, and amateur observers a straightforward 
tool to visualize double star data with minimal effort. The toolkit can be run 
locally (via \texttt{python scriptname.py}) or easily pasted into Google Colab 
notebooks for immediate, interactive plotting.

\section{Overview}

\textbf{Motivation and Scope.} Many double star research projects gather data from 
multiple sources:
\begin{itemize}
  \item \textbf{Historical observations}: Typically spanning decades or centuries, 
        measured in either polar coordinates (P.A., Sep) or Cartesian coordinates (X, Y).
  \item \textbf{Gaia DR3}: High-precision astrometric data, often in polar form or 
        convertible to Cartesian.
  \item \textbf{LCO measurements}: Modern measurements from the Las Cumbres Observatory 
        telescope network.
  \item \textbf{Predictions / Theoretical points}: Coordinates derived from orbital 
        fits (e.g., \cite{Izmailov2019}).
\end{itemize}

\noindent
\textbf{Key Features.}
\begin{itemize}
  \item One color-coded scatter for historical data (based on date).
  \item Separate styling for Gaia measurements (e.g., red circles).
  \item Averaging of multiple LCO points into one marker (green “X”).
  \item Optional predicted or theoretical data, plotted as a distinct color/marker 
        (light-blue “X”).
  \item A plot with matching numeric ranges on the X- and Y-axes to avoid distortion.
\end{itemize}

\section{Installation}

The scripts require Python 3.7 or higher. Required packages include:
\begin{lstlisting}[language=bash]
pip install numpy matplotlib
\end{lstlisting}

\noindent
You can find the complete repository of scripts at:
\begin{center}
\url{https://github.com/xinyue0221/double-star-plot-scripts}
\end{center}

\noindent
To use these scripts, you can:
\begin{enumerate}
  \item \textbf{Clone or download the repository} (e.g., via Git or as a ZIP file).
  \item \textbf{Navigate to the directory} where the scripts are located.
  \item \textbf{Run the scripts directly} or import their functions into your own 
        Python scripts or Jupyter/Colab notebooks.
\end{enumerate}

\section{Usage}

These scripts are intended to be as simple as possible. In many cases, you only need 
to provide arrays of coordinates (historical data, Gaia data, LCO data) and a list 
of observation dates. The script handles color-coding, marker types, axis scaling, 
and legends automatically. 

\subsection{Running Scripts Standalone}

Each script has an \texttt{if \_\_name\_\_ == "\_\_main\_\_"} block containing a 
working example. For instance, to test the first script:

\begin{lstlisting}[language=bash]
python script_1_plot_double_star_three_datasets_average_lco.py
\end{lstlisting}

This will pop up a \texttt{matplotlib} window displaying a sample double star plot.

\subsection{Using in Jupyter or Google Colab}

You can copy and paste the function definitions from these scripts into a Jupyter 
notebook or Google Colab cell. Then simply call:

\begin{lstlisting}[language=Python]
from script_1_plot_double_star_three_datasets_average_lco import (
    plot_double_star_three_datasets_average_lco
)

# Example data
hist_dates = [1900, 1950, 2000, 2020]  # hypothetical
hist_PAs   = [0, 45, 90, 135]         # in degrees
hist_Seps  = [10,  9.5, 9, 8.5]

plot_double_star_three_datasets_average_lco(
    hist_dates, hist_PAs, hist_Seps,
    gaia_PAs=None, gaia_Seps=None,
    lco_PAs=None, lco_Seps=None
)
\end{lstlisting}

The script will generate a figure in Colab with a 1:1 aspect ratio.

\section{Scripts}

This toolkit contains three primary Python scripts, each tailored to a specific 
coordinate system or use case.

\subsection{Script 1: \texttt{plot\_double\_star\_three\_datasets\_average\_lco.py}}

\begin{lstlisting}[language=Python]
plot_double_star_three_datasets_average_lco(
    hist_dates, hist_PAs, hist_Seps,
    gaia_PAs=None, gaia_Seps=None,
    lco_PAs=None, lco_Seps=None,
    gaia_label="Gaia DR3 measurement",
    lco_label="[TIME] LCO measurement (average)"
)
\end{lstlisting}

\paragraph{Purpose}
When your historical, Gaia, and LCO measurements are all in polar coordinates 
(P.A., Sep), this function:

\begin{itemize}
  \item Converts all polar data to $(x, y)$ under the convention:
  \[
    x = \text{Sep} \cdot \sin(\text{P.A.}), \quad
    y = -\text{Sep} \cdot \cos(\text{P.A.})
  \]
  where P.A. is in degrees measured from North ($0^\circ$) toward East ($90^\circ$).
  \item Plots historical data color-coded by date (older = lighter, newer = darker).
  \item Plots Gaia data as red circles, if provided.
  \item Averages LCO data to a single green ``X.''
  \item Enforces the same numeric range on both axes for an undistorted view.
\end{itemize}

\subsection{Script 2: \texttt{plot\_double\_star\_three\_datasets\_XY.py}}

\begin{lstlisting}[language=Python]
plot_double_star_three_datasets_XY(
    hist_x, hist_y, hist_dates,
    gaia_x=None, gaia_y=None,
    lco_x=None, lco_y=None,
    gaia_label="Gaia DR3 measurement",
    lco_label="LCO measurement (average)"
)
\end{lstlisting}

\paragraph{Purpose}
For historical, Gaia, and LCO data already in $(x, y)$ form, this script:
\begin{itemize}
  \item Plots historical $(x,y)$ data color-coded by date.
  \item Plots Gaia data as red circles, if provided.
  \item Averages LCO data to a single green ``X,'' if multiple points exist.
  \item Applies a 10\% margin around the data to give some breathing room.
  \item Sets a 1:1 aspect ratio to preserve geometry.
\end{itemize}

\subsection{Script 3: \texttt{plot\_double\_star\_with\_prediction.py}}

\begin{lstlisting}[language=Python]
plot_double_star_with_prediction(
    hist_x, hist_y, hist_dates,
    gaia_x=None, gaia_y=None,
    gaia_label="Gaia DR3 measurement",
    lco_x=None, lco_y=None,
    lco_label="LCO measurement (average)",
    pred_x=None, pred_y=None,
    pred_label="Prediction"
)
\end{lstlisting}

\paragraph{Purpose}
When you have $(x, y)$ coordinates for historical data, Gaia, LCO, and an additional 
predicted point (for example, from orbit solutions), this function:

\begin{itemize}
  \item Color-codes historical points by observation date.
  \item Marks Gaia data as red circles.
  \item Averages LCO data into one green ``X.''
  \item Plots a predicted coordinate as a light-blue ``X,'' if provided.
  \item Ensures a 1:1 aspect ratio and a clear legend.
\end{itemize}

\section{Examples}

Each script contains an example code block under the 
\texttt{if \_\_name\_\_ == "\_\_main\_\_"} guard, which you can run directly to see 
a sample plot. Alternatively, you can import the functions into your own scripts or 
notebooks.

\begin{figure}[h]
\centering
\includegraphics[width=0.45\textwidth]{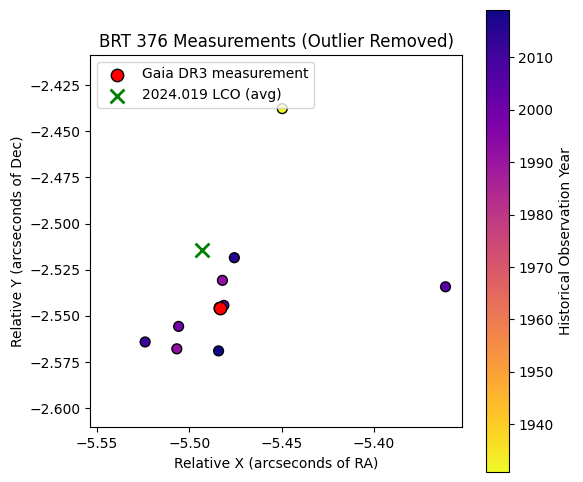}
\caption{Example output of a double star plot using \texttt{plot\_double\_star\_three\_datasets\_XY.py}. 
Historical data is color-mapped by date, Gaia data is shown in red, and LCO data 
is averaged to one green ``X.'' (Fictitious data for demonstration.)}
\label{fig:example}
\end{figure}

\noindent
\textbf{Title and Axis Labels.} You can customize the plot by adding or editing:
\begin{lstlisting}[language=Python]
ax.set_title("My Double Star Plot Title")
ax.set_xlabel("Delta X (arcseconds)")
ax.set_ylabel("Delta Y (arcseconds)")
\end{lstlisting}

\noindent
\textbf{Colormap.} By default, the script uses \texttt{plasma\_r} for color-coding 
historical data. You can replace it with any \texttt{matplotlib} colormap, such as 
\texttt{viridis}, \texttt{coolwarm}, etc.

\section{License}

\noindent
\textbf{MIT License} \textcopyright\ Xinyue Wang

\medskip
Permission is hereby granted, free of charge, to any person obtaining a copy of 
this software and associated documentation files (the ``Software''), to deal in the 
Software without restriction, including without limitation the rights to use, copy, 
modify, merge, publish, distribute, sublicense, and/or sell copies of the Software, 
and to permit persons to whom the Software is furnished to do so, subject to the 
following conditions:

\begin{quote}
\small
The above copyright notice and this permission notice shall be included in all 
copies or substantial portions of the Software.

THE SOFTWARE IS PROVIDED ``AS IS'', WITHOUT WARRANTY OF ANY KIND, EXPRESS OR IMPLIED, 
INCLUDING BUT NOT LIMITED TO THE WARRANTIES OF MERCHANTABILITY, FITNESS FOR A 
PARTICULAR PURPOSE AND NONINFRINGEMENT. IN NO EVENT SHALL THE AUTHORS OR COPYRIGHT 
HOLDERS BE LIABLE FOR ANY CLAIM, DAMAGES OR OTHER LIABILITY, WHETHER IN AN ACTION 
OF CONTRACT, TORT OR OTHERWISE, ARISING FROM, OUT OF OR IN CONNECTION WITH THE 
SOFTWARE OR THE USE OR OTHER DEALINGS IN THE SOFTWARE.
\end{quote}

\section{Citation and Related Work}

If you use these scripts in your research, please cite this toolkit. An example 
citation for the present work might be:

\begin{quote}
\small
X. Wang, \emph{A Python Toolkit for Plotting Double Star Observations with 1:1 Aspect Ratio}, 2025,\\
\url{https://doi.org/10.48550/arXiv.2502.13340}
\end{quote}

These scripts have been used in:
\begin{itemize}
  \item \textbf{Wang, Xinyue (2025).} \textit{Astrometric Measurements and Analysis of 
        Double Star System BRT 376.} 
        \href{https://doi.org/10.48550/arXiv.2502.11648}{10.48550/arXiv.2502.11648.} \cite{Wang2025}
  \item \textbf{Janakiraman, R., Li, S., Humphreys, M.S.M., Arruda, M., Wang, X., 
        \& Tock, K.} (unpublished, submitted to the \emph{Journal of Double Star 
        Observations}). \emph{Astrometric Measurements of 05276--0843 HLD 75, 
        06347--8239 WFC 38, 05445--2058 KPP 125, 10296+3757 HJ 2532 AB and 
        09174+2339 STF 1332 (HD 79872)}.
\end{itemize}

In these works, the toolkit facilitated the visualization of astrometric data from 
historical, Gaia, and LCO observations, simplifying cross-comparison and ensuring 
accurate depiction of stellar separation and position angles.

\section{Conclusion}

The Python scripts described here aim to simplify the visualization of double star 
astrometric data from multiple sources (historical catalogs, Gaia DR3, LCO, and 
theoretical predictions). By enforcing a square, 1:1 aspect ratio, these plots 
avoid misleading distortions and more faithfully represent the true geometry of 
the system. Easy integration with Google Colab and other Python environments allows 
for rapid adoption by professional astronomers, students, and citizen scientists 
alike.

Future enhancements might include:
\begin{itemize}
  \item Automatic unit conversion and consistency checks (e.g., arcseconds vs. 
        milliarcseconds).
  \item Interactive plots using \texttt{ipywidgets} or \texttt{plotly}.
  \item Expanded color-map customization and advanced plotting themes.
\end{itemize}

Ultimately, this toolkit should help foster more intuitive presentations of 
astrometric data, empowering researchers to glean more accurate insights into double 
star orbits and evolution.

\section*{Acknowledgments}
The idea for this toolkit originated during a research seminar course at Stanford 
University Online High School. Special thanks to the instructors and peers for 
their feedback and suggestions. The author also appreciates the open-source 
community for providing many of the libraries that made this toolkit feasible.


\end{document}